# Alignment-dependent fluorescence emission induced by tunnel ionization of carbon dioxide from lower-lying orbitals

Jinping Yao[1], Guihua Li[1,7], Xinyan Jia[2], Xiaolei Hao[3,4], Bin Zeng[1], Chenrui Jing[1,7], Wei Chu[1], Jielei Ni[1,7], Haisu Zhang[1,7], Hongqiang Xie[1,7], Chaojin Zhang[1], Zengxiu Zhao[5], Jing Chen[3,4,†], Xiaojun Liu[6,§], Ya Cheng[1,*] and Zhizhan Xu[1,#]

[1] State Key Laboratory of High Field Laser Physics, Shanghai Institute of Optics and Fine Mechanics, Chinese Academy of Sciences, Shanghai 201800, China,
[2] Quantum Optoelectronics Laboratory, Southwest Jiaotong University, Chengdu 610031, China
[3] Key Laboratory of High Energy Density Physics Simulation, Center for Applied Physics and Technology, Peking University, Beijing 100084, China
[4] Institute of Applied Physics and Computational Mathematics, Beijing 100088, China
[5] Department of Physics, National University of Defense Technology, Changsha 410073, China
[6] State Key Laboratory of Magnetic Resonance and Atomic and Molecular Physics, Wuhan Institute of Physics and Mathematics, Chinese Academy of Sciences, Wuhan 430071, China
[7] University of Chinese Academy of Sciences, Beijing 100049, China

[†]*chen_jing@iapcm.ac.cn*

[§]*xjliu@wipm.ac.cn*

[*]*ya.cheng@siom.ac.cn*

[#] *zzxu@mail.shcnc.ac.cn*

**Abstract**

Study on ionization process of molecules in an intense infrared laser field is of paramount interest in strong-field physics and constitutes the foundation of imaging of molecular valence orbitals and attosecond science. We show measurement of alignment-dependent ionization probabilities of the lower-lying orbitals of the molecules by experimentally detecting alignment-dependence of fluorescence emission from tunnel ionized carbon dioxide molecules. The experimental measurements are compared with the theoretical calculations of strong field approximation (SFA) and molecular ADK models. Our results demonstrate the feasibility of an all-optical approach for probing the ionization dynamics of lower-lying orbitals of molecules, which is still difficult to achieve until now by other techniques. Moreover, the deviation between the experimental and theoretical results indicates the incompleteness of current theoretical models for describing strong field ionization of molecules.

Tunnel ionization is a fundamental mechanism underlying the extreme nonlinear strong field atomic and molecular phenomena such as high-order harmonic generation (HHG), above-threshold ionization (ATI), etc [1-5]. The characteristic of tunnel ionization is that it depends exponentially on the strength of light field and inversely exponentially on the ionization potential of the target atoms. Due to its extreme sensitivity on the strength of light field, tunnel ionization typically occurs in a small fraction of the period of light field at the crest of the light field, laying the foundation of attosecond science and technology [6]. On the other hand, the sensitivities of tunnel ionization on the ionization potential as well as the orbital structures and symmetries render itself an efficient probe of the electronic structures of molecules in either a static or a dynamic manner [7,8]. Based on this expectation, considerable efforts have been devoted to unlocking the geometric information of the highest occupied molecular orbitals (HOMOs) of molecules from their alignment-dependent tunnel ionization rates [9-11]. Indeed, for some molecules such as $N_2$ and $O_2$, satisfying agreement between the experimental and theoretical results has been obtained. However, when examining the carbon dioxide ($CO_2$) molecule, the alignment-dependent ionization rate of its HOMO orbital calculated using the molecular Ammosov-Delone-Krainov (MO-ADK) theory fails to reproduce the experimental observation [9]. Moreover, a recent study on wavelength-dependence of tunnel ionization of $O_2$ in strong laser fields confirms that interference between the electron wavepackets ionized from different cores, which is not included in the MO-ADK theory, actually plays an important role [12]. These facts provide clear evidences on the complexity of ionization mechanisms of molecules in strong laser fields, whose complete understanding therefore requires development of new experimental and theoretical approaches such as the one described below in this Letter.

Although experimental investigations of tunnel ionization of molecules are frequently carried out based on HHG or ATI techniques, they usually lack the capability to directly access the lower-lying orbitals of molecules due to the difficulties in distinguishing the contributions to the total ionization rate from the individual orbitals. On the other hand, it has recently been discovered that for some molecules, their lower-lying orbitals can substantially contribute to the ionization process because their ionization potentials are close to that of the outmost orbital and, more importantly, have different alignment dependence of ionization probabilities from that of the outmost orbital [13]. Therefore, establishment of a technique that allows direct observation of tunnel ionization from individual lower-lying orbitals will undoubtedly be of great importance. In this Letter, we show that such an ambition can be realized by employing the fact that the intensities of the fluorescence lines directly reflect the ionization probabilities from lower-lying molecular orbitals [14,15]. Particularly, we have recently observed that tunnel ionization from lower-lying molecular orbitals can produce not only molecular ions at excited states, but also population inversion in the ionized molecules by which lasing can be initiated [14]. Thus, we expect that by deconvolution of the measured alignment dependence of the fluorescence, we can obtain the alignment dependence of the ionization probability of HOMO-1 and HOMO-2 orbitals of molecules. In particular, $CO_2$ molecule was chosen in this experiment for the technical reason that the fluorescence emissions resulted from the ionization of $CO_2$ from its HOMO-1 and HOMO-2 orbitals are both in the visible range. Physically speaking, investigation of the lower-lying orbitals of $CO_2$ will help clarifying its ionization mechanism in the tunnel regime (i. e., the Keldysh parameter $\gamma$ less than or close to 1) which has been a long-standing controversial issue and a topic of hot debate, too [16].

In our alignment-dependent fluorescence experiment, the pump and probe pulses were obtained by splitting linearly polarized output laser pulses (800 nm, ~40 fs, 1 kHz, 6 mJ) from a Ti:sapphire laser system (Legend Elite-Duo, Coherent Inc.) using a 50% beam splitter. The pump beam, which was used for aligning the molecules, passed through an inverted telescope for reducing its beam diameter by half. Then, the two beams were recombined by a 50% beam splitter and collinearly focused by a position-adjustable lens into a chamber filled with $CO_2$ gas. In order to gain a high degree of alignment at the room temperature, the pump beam was slightly broadened to ~60 fs, whereas the pulse duration of probe beam was maintained at ~40 fs. Because of its smaller beam size, the peak intensity of the pump beam at the focal spot is significantly lower than that of probe beam, as we attempt to minimize the fluorescence signal (i. e., the ionization) produced by the pump pulse alone. Meanwhile, special care was taken to choose a relatively low gas pressure of ~2 mbar to minimize both the intensity change of the probe pulse and the variation of the fluorescence spectrum caused by the spatiotemporal modulation of the refractive index in aligned molecules [17,18]. Under such a low gas pressure condition, the plasma defocusing effect is very weak and can be safely ignored as we have confirmed in our experiment. A delay line and a half-wave plate were inserted into the pump-beam path for adjusting the time delay and the relative angle between the polarization axes of the pump and probe pulses, respectively. To avoid the change of pump intensity caused by the rotation of the half-wave plate, the splitter used to combine two beams is arranged at a small angle. The fluorescence spectra of $CO_2^+$ were collected from the backward direction (i.e., opposite to the direction of laser propagation) using a fused silica lens and then were detected by a grating spectrometer (Shamrock 303i, Andor). With this scheme, the alignment dependence of spontaneous emission, which leads to anisotropic fluorescence emission around the optical axis of the laser beam, could be minimized, and thus the measurement of fluorescence signal as a function of molecular alignment angle is equivalent to measurement of the alignment-dependent ionization probability of molecules from their corresponding lower-lying orbitals. It should be stressed that in our experiment, the peak intensity of the probe pulses at the focus was set at $\sim 4 \times 10^{14}$ W/cm$^2$, resulting in a Keldysh parameter $\gamma \approx 0.6$ calculated using the ionization potential of HOMO-1 orbital of $CO_2$. This ensures that the ionization processes investigated here all occurs in the tunnel regime.

From the electronic configuration of neutral $CO_2$ molecules $(1\sigma_g)^2(1\sigma_u)^2(2\sigma_g)^2(3\sigma_g)^2(2\sigma_u)^2(4\sigma_g)^2(3\sigma_u)^2(1\pi_u)^4(1\pi_g)^4$, we can know that removal of electron from HOMO-1 (i. e., $1\pi_u$ orbital) and HOMO-2 (i. e., $3\sigma_u$ orbital) will produce $CO_2^+$ in the first excited state ($A^2\Pi_u$) and the second excited state ($B^2\Sigma_u$), respectively. Decay from these excited states to the ground state ($X^2\Pi_g$) will lead to spontaneous fluorescence emissions at 338 nm ($A^2\Pi_u \rightarrow X^2\Pi_g$, 1→0) and 289 nm ($B^2\Sigma_u \rightarrow X^2\Pi_g$, 0→0), as illustrated in Fig. 1(a). A typical fluorescence spectrum from the unaligned $CO_2$ molecules measured by the grating spectrometer is shown in Fig. 1(b). Throughout the experiment, a 1200 grooves/mm grating and a 300 μm slit were used, and thus the spectral resolution in the range of 280-350 nm was ~0.07 nm. In fact, transitions between different vibrational levels from the excited state to the ground state of $CO_2^+$ correspond to multiple fluorescence peaks. In this work, we only choose two typical fluorescence lines at wavelengths of 338 nm and 289 nm for investigating the alignment dependence of ionizations of $CO_2$ from the HOMO-1 and HOMO-2 orbitals, respectively. Accordingly, the wave functions of HOMO-1 and HOMO-2 orbitals of $CO_2$ molecules, which were calculated using the GAMESS code [19], are illustrated in Figs. 1(c) and (d), respectively.

Figures 2(a) and (b) show the measured fluorescence signals at the respective wavelengths of 338 nm and 289 nm as functions of the time delay between the pump and probe pulses. To augment the details in the variation of fluorescence signal with the changing time delay, the time range in Fig. 2 is confined to half of the rotational period $T_{rot}$ ($T_{rot} \approx$ 42.7ps). Here, the zero time delay ($t=0$), which corresponds to the best temporal overlap between the pump and probe pulses, was determined by maximizing the white light generation in air. To achieve a high degree of alignment of the $CO_2$ molecules and keep the background fluorescence signal as weak as possible, the beam diameter as well as the intensity of the pump pulse were carefully optimized. The polarization directions of the pump and probe pulses are set to be parallel to each other. Figure 2(c) shows the calculated temporal evolution of the degree of alignment $\langle \cos^2 \theta \rangle$ of $CO_2$ molecules, which is defined as $\langle \cos^2 \theta \rangle \equiv \sum_{J_0=0}^{\infty} g_{J_0} \sum_{M_0=-J_0}^{J_0} \langle \cos^2 \theta \rangle_{J_0, M_0}$ by assuming an initial Boltzmann distribution of the molecular ensemble. Here, $\theta$ is the angle between the molecular axis and the polarization direction of pump pulses. Under this condition, we have $\langle \cos^2 \theta \rangle_{J_0, M_0} = \langle \Psi(t) | \cos^2 \theta | \Psi(t) \rangle$, and $g_{J_0}$ the Boltzmann weights of the initial state $|J_0, M_0\rangle$. For the nonadiabatic field-free alignment, the rotational wave packet $\Psi(t)$ is calculated by solving time-dependent Schrödinger equation (TDSE) based on the rigid rotor model [20,21]. In our simulation, the parameters of alignment pulses (i.e., the pump pulses) were the same as the experimental ones (i.e., 800 nm, 60 fs, $1 \times 10^{14}$ W/cm$^2$, 300 K). In Fig. 2(c), it can be seen that at $t_1 = 21.25$ps, the molecule ensemble is aligned in the direction parallel to the polarization of pump pulses, whereas at $t_2 = 21.58$ps, an anti-aligned ensemble is generated, namely, the molecular axis is perpendicular to the polarization of pump pulses. Apparently, the experimental observations shown in Figs. 2(a) and (b), i.e., the changes of the fluorescence intensities with the time delay, are highly relevant to the theoretical curve in Fig. 2(c). The evolution of fluorescence signal at 289 nm follows the same trend of the theoretical curve, whereas the evolution of fluorescence signal at 338 nm shows an opposite dependence on the time-dependent alignment degree. This finding reflects the difference in the alignment dependence of ionization rates of $CO_2$ from the HOMO-1 and HOMO-2 orbitals. Since the fluorescence at 338 nm originates from the HOMO-1 orbital with a $\pi_u$ symmetry, this signal becomes the weakest when the parallel alignment (i.e., the molecular axis is parallel to the polarization direction of the probe pulses) was achieved. In contrast, the 289 nm signal originates from the HOMO-2 orbital which has a $\sigma_u$ symmetry, thus it reached its maximum under the parallel alignment condition. In addition, it is noteworthy that the fluorescence signal at 289 nm wavelength reaches its maximum and minimum at the time delay of 21.27 ps and 21.57 ps, respectively. In contrast, the situation for the fluorescence signal at 338 nm is completely opposite. Below we will show that by measuring the fluorescence signal as a function of the alignment angle, the alignment dependence of the ionization probability from an individual lower-lying orbital can be directly examined without interference from others, enabling measurement of alignment-dependent ionization probabilities of lower-lying orbitals of molecules.

As indicated by the black dashed line in Fig. 2, the parallel alignment of the ensemble of $CO_2$ molecules was achieved at the time delay of 21.27 ps. The alignment dependence of fluorescence intensity can thus be examined by varying the relative angle $\alpha$ between the pump polarization and the probe polarization at this fixed time delay. The relative angle $\alpha$ was gradually increased by rotating the half-wave plate in the pump-beam path with an angular step value of 4° when the polarization axis of probe pulses was maintained in the horizontal plane. Figures 3(a) and (b) present the measured fluorescence intensities at the wavelengths of 338 nm and 289 nm, respectively, as a function of the relative angle $\alpha$ between the polarization axes of the pump and the probe beams. It should be mentioned that in Figs. 3(a) and (b), a weak fluorescence background generated by the pump pulses (i. e., the alignment beam) alone, has been subtracted from the total fluorescence signal. Monotonic increase of the fluorescence signal at 338 nm is observed for an angular range from 0° to 90°; whereas the signal at 289 nm exhibits a different behavior, i. e., it decreases with the increasing the angle $\alpha$. The mechanism underlying the different behaviors is the sensitivity of tunnel ionization on the geometries of molecular orbitals.

Owing to the fact that perfect molecular alignment is practically impossible, strictly speaking, the measured alignment-dependent fluorescence intensity $S_{exp}(\alpha, t = 21.27\text{ps})$ in Fig. 3 can be regarded as the convolution of the alignment-dependent ionization probability $P_{ion}(\theta')$ and the time-dependent distribution of aligned molecular axis $\rho(\theta, t_1)$ [7,9], thus it can be expressed as:

$$S_{cal}(\alpha, t_1) = \int_{\varphi'=0}^{2\pi} \int_{\theta'=0}^{\pi} \rho(\theta(\theta', \varphi', \alpha), t_1) P_{ion}(\theta') \cdot \sin\theta' d\theta' d\varphi' \quad (1)$$

Here, $\rho(\theta, t_1) = \sum_{J_0=0}^{\infty} g_{J_0} \sum_{M_0=-J_0}^{J_0} \langle \Psi(t) | \Psi(t) \rangle \big|_{t=t_1}$ is calculated by solving the TDSE at the moment of the first half-revival ($t_1 = 21.25\text{ps}$) with the same parameters as used in

Fig. 2(c). The time-dependent distribution $\rho(\theta, t_1)$ in the frame of pump pulses can be transformed to the frame of probe pulses using the relation $\cos\theta = \cos\alpha \cos\theta' + \sin\alpha \sin\theta' \cos\varphi'$, where $\theta'$ and $\varphi'$ are the polar and azimuthal angles in the frame of probe pulses, respectively. To retrieve the alignment-dependent ionization probability $P_{ion}(\theta')$, we first expand it into the form $P_{ion}(\theta') = \sum_{n=0}^{3} C_n \cos(2n\theta')$, and then determine the coefficients $C_n$ by fitting the calculated $S_{cal}(\alpha, t_1)$ to the measured $S_{exp}(\alpha, t = 21.27\text{ps})$. As indicated by the red solid lines in Figs. 3(a) and (b), the calculated fluorescence signal $S_{cal}(\alpha, t_1)$ are in good agreement with the measured curves.

Generally, besides *ab initio* methods [10], two analytical models, i.e., the strong-field-approximation (SFA) theory [22-24] and MO-ADK theory [3], have been widely applied to describe the molecular ionization in strong laser fields. However, the underlying physical pictures of the ionization process are intrinsically different in these two models. The MO-ADK theory treats the ionization of molecules as a tunneling process which is essentially identical to that of atoms. Hence the ionization is mainly determined by the asymptotic behavior of the bound state wave function at large distance [3]. However, in the SFA theory, the ionization is calculated by the S-matrix theory which treats the external laser field in a nonperturbative way. For molecules like $O_2$, the interference between ionized wave packets emitted from different nuclei in the molecule is believed to play an essential role in the ionization process [12,22]. So far, no consensus has been achieved for the underlying mechanism of the characteristic molecular ionization. On the other hand, study of alignment-dependent total ionization probability of $CO_2$ with ion measurement shows significant variance from the prediction of the MO-ADK theory [9] and leads to intensive investigations [10,25,26].

To shed more light on this controversial problem and check the validity of our experiment method, we compare the retrieved alignment-dependent ionization probabilities from both the HOMO-1 and HOMO-2 orbitals with simulation results of the length-gauge SFA theory and MO-ADK theory in Fig. 4. In these calculations, ionization rates of electrons from the HOMO-1 and HOMO-2 orbitals are evaluated using formula in Refs. [23,24] (for length-gauge SFA) and Ref. [3] (for MO-ADK). The wavefunctions are obtained by GAMESS code, which are shown in Figs. 1(c), (d). The good agreement between the results obtained by alignment-dependent fluorescence measurement and SFA simulation provides a strong evidence on the effectiveness of our technique for directly probing the ionization dynamics of lower-lying orbitals. For the HOMO-1 orbital in Fig. 4(a), the deconvoluted $P_{ion}(\theta')$, which gives rise to the fluorescence signal at 338 nm, shows a heart-shaped distribution with the ionization probability peaked at ~65 degrees, which is consistent with the prediction of the SFA calculation. It is intriguing that similar to the alignment-dependent total ionization probability of $CO_2$ shown in Ref. [9], our results significantly differs from that given by the MO-ADK theory. Moreover, it is noteworthy that the laser parameters used in our experiment ensure that the ionization happens in the tunneling regime ($\gamma\approx 0.6$ for HOMO-1), different from that used in Ref. [9] which is about $\gamma\approx 1$ and multiphoton effect may affect the ionization process. The MO-ADK calculation for HOMO-1 orbital shows that the alignment-dependent ionization probability peaks at ~30 degrees as shown in Fig. 4(a), which coincides with the peak of the angular distribution of the asymptotic electron density of $CO_2$. For the HOMO-2 orbital in Fig. 4(b), the deconvoluted $P_{ion}(\theta')$ (retrieved using the fluorescence signal at 289 nm), gradually decreases from 0° to 90°, which qualitatively follows the theoretical curves obtained by both length-gauge SFA and MO-ADK calculations. However, as compared to HOMO-1 orbital in Fig. 4(a), a more significant difference between the SFA and experimental results can be found for HOMO-2 orbital, which may be partially attributed to an intramolecular coupling between the $A^2\Pi_u$ state and excited vibronic levels of the $B^2\Sigma_u$ state [27] or inaccuracy of theoretical calculation. In addition, the apparent inconsistency between the experimental measurements and the MO-ADK simulations for the HOMO-1 orbital indicates that the MO-ADK theory, which considers the ionization as a pure tunneling process and has been widely accepted for describing ionization of atoms, may not be able to accurately describe the ionization process of molecules. In contrast, the SFA theory well reproduces the experimental results for both HOMO-1 and HOMO-2 orbitals, implying that the interference effect, which is inherently accounted for in the SFA theory, should play an essential role in the molecular ionization. More detailed analysis is ongoing and will be presented elsewhere.

In conclusion, we have shown that the fluorescence emissions from tunnel ionized molecules are sensitive to the molecular alignment angles, which provides an efficient way to investigation of the alignment-dependent ionization from the lower-lying orbitals. As an all optical means concerning only the wavelengths in the visible and near ultraviolet regions, the technique is simple and cost-effective. Although only $CO_2$ molecule is investigated in this work as a model molecule, in principle, this technique can be applied to a wide range of molecules, as far as their fluorescence emissions are in the suitable spectral ranges and detectable by the available spectrometers. For example, other than $N_2$, molecules such as $O_2$, HCl, $H_2O$, etc. can all be potential candidates. Due to the exceptionally rich physics in the ionization process of molecules, the unique capability of directly probing the ionization from the lower-lying orbitals in an angular-resolvable manner will provide us not only the potential of imaging of lower-lying molecular orbitals, but also an ideal testbed for existing theoretical models. The new insight gained by comparing our experimental results with SFA and MO-ADK calculations has already provided a good example on this aspect.

This work is supported by the National Basic Research Program of China (Grant No. 2011CB808100 and No. 2013CB922200), and National Natural Science Foundation of China (Grant Nos. 11127901, 11134010，60921004, 11074026, 10925420, 61275205, 11204332, and 11274050).

**Captions of figures:**

Fig. 1 (Color online) (a) Energy level diagram of ionized and neutral $CO_2$ molecules. The fluorescence emissions at wavelengths of 338 nm and 289 nm originate from the transitions ($A^2\Pi_u \rightarrow X^2\Pi_g$, $1\rightarrow 0$) and ($B^2\Sigma_u \rightarrow X^2\Pi_g$, $0\rightarrow 0$). (b) A typical fluorescence spectrum measured using a grating spectrometer. Calculated electron wavepackets of lower-lying orbitals of $CO_2$ by GAMESS code: (c) HOMO-1, and (d) HOMO-2.

Fig. 2 (Color online) Temporal evolution of the fluorescence signals at (a) 289 nm and (b) 338 nm wavelengths from aligned $CO_2$ molecules at the moment around the first half-revival. (c) The calculated time-dependent alignment parameter $\langle \cos^2 \theta \rangle$.

Fig. 3 (Color online) Fluorescence signals at (a) 338 nm and (b) 289 nm as functions of the relative polarization angle $\alpha$ between pump and probe pulses at the time delay of 21.27 ps. The error bars in the fluorescence signals are estimated from multiple measurements. Red solid lines are corresponding theoretical fitting curves.

Fig. 4 (Color online) Experimentally retrieved alignment-dependent ionization probability $P_{ion}(\theta')$ of $CO_2$ molecules (blue solid lines) for (a) HOMO-1 and (b) HOMO-2 orbitals. Corresponding results obtained using the length-gauge SFA and the MO-ADK calculations are indicated by the red dot-dashed lines and the green dotted lines, respectively.

**Fig. 1**

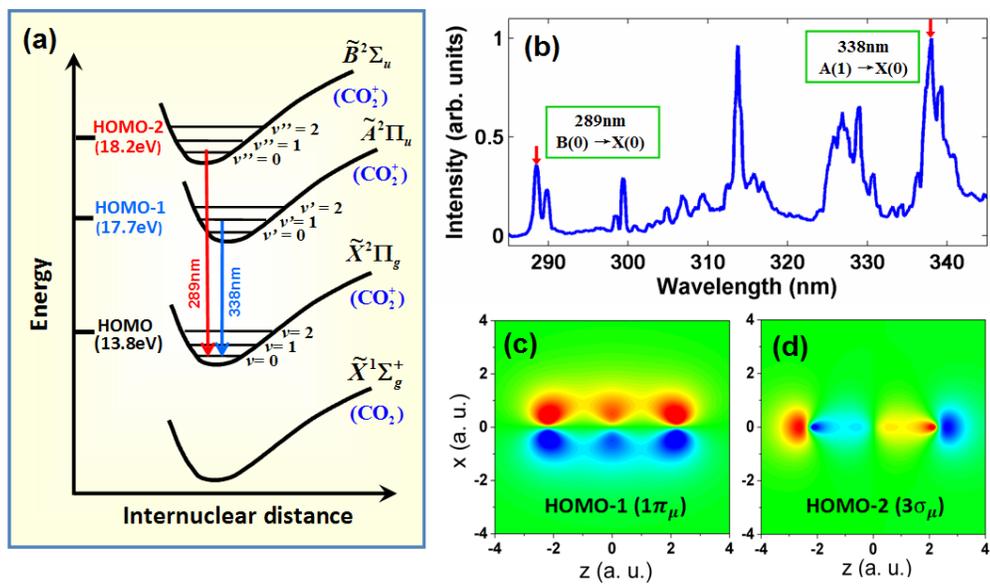

**Fig. 2**

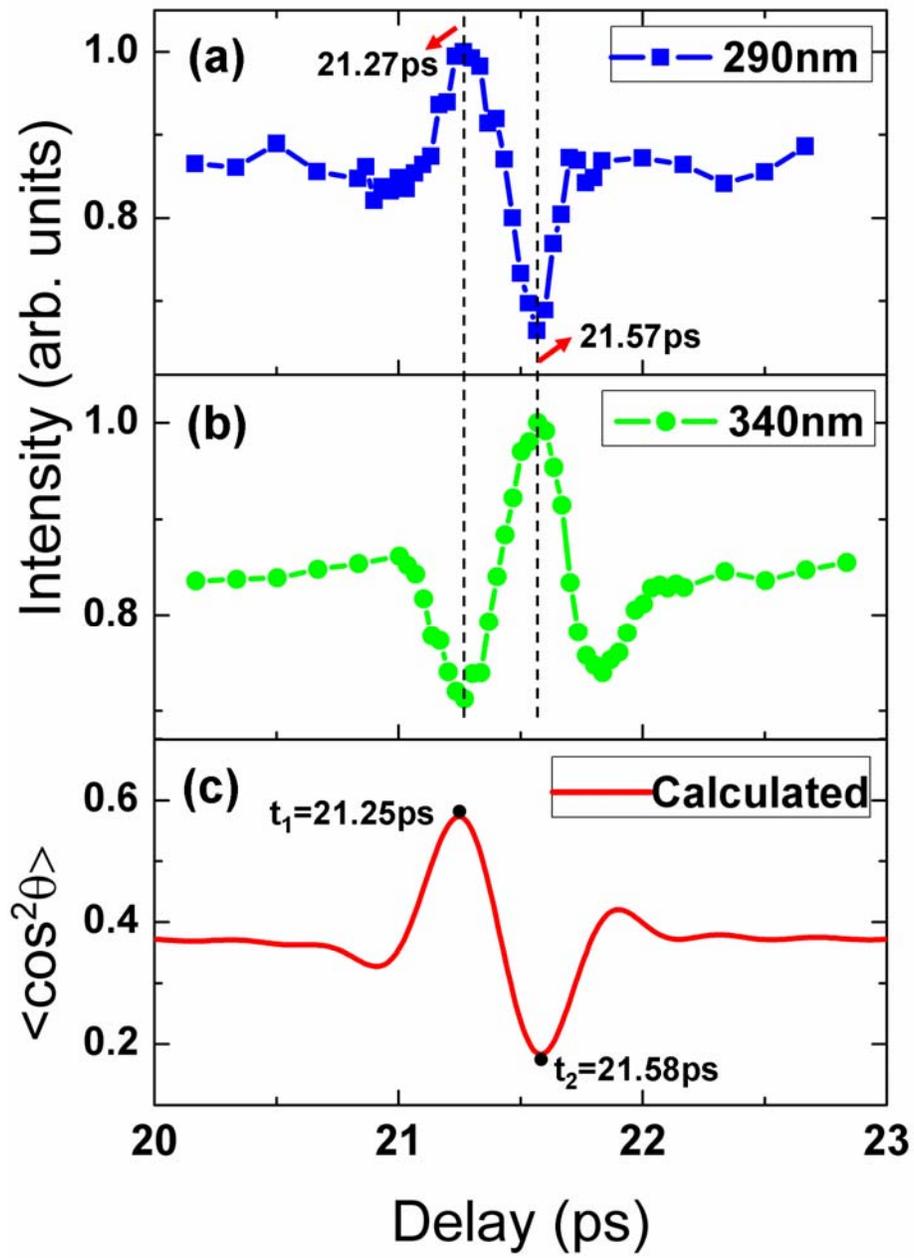

**Fig. 3**

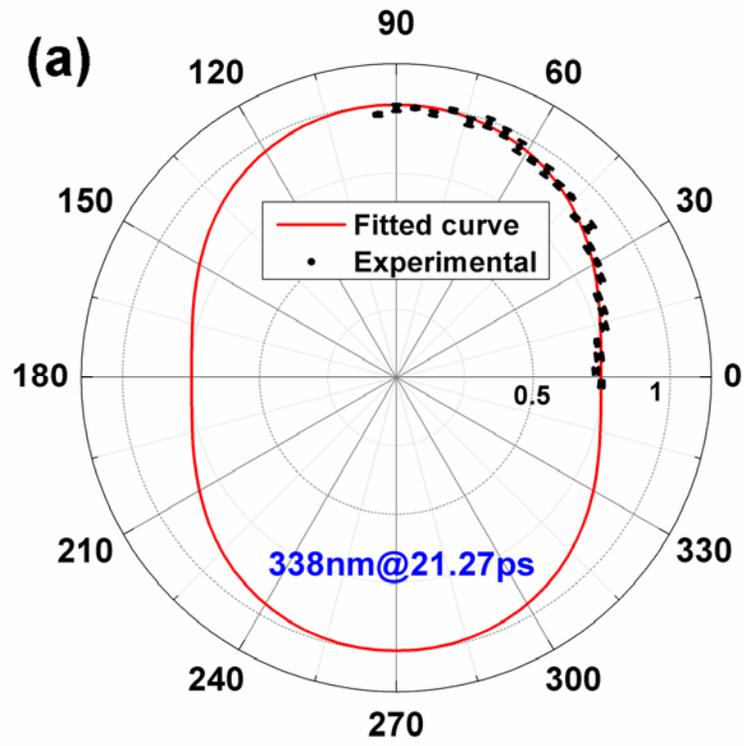

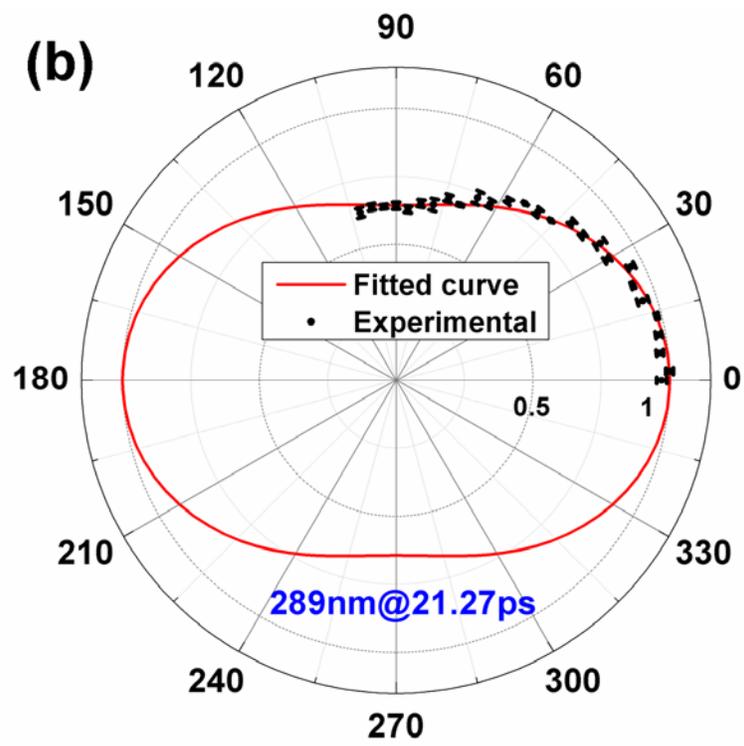

**Fig. 4**

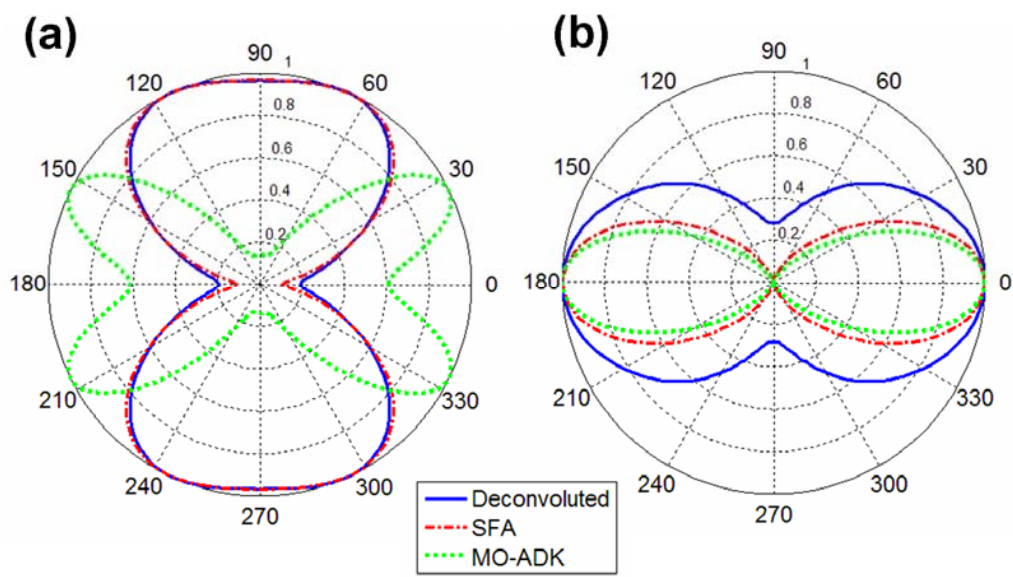